\documentclass{aa}
\begin{document}

   \title{Effects of disc asymmetries on astrometric measurements}

   \subtitle{Can they mimic planets?}

   \author {Q. Kral\inst{1}
     \and 
          J. Schneider\inst{2}%\fnmsep\thanks{Just to show the usage
  \and
	  G. Kennedy\inst{1}%\fnmsep\thanks{Just to show the usage
	\and 
       D. Souami\inst{3} }

   \offprints{Q. Kral, J. Schneider}

   \institute{
		$(1)$ Institute of Astronomy, University of Cambridge, Madingley Road, Cambridge CB3 0HA, UK\\
   		$(2)$ Observatoire de Paris, LUTh-CNRS, UMR 8102, 92190, Meudon, France\\
		$(3)$ NaXys, University of Namur, Rempart de la Vierge 8, 5000 Namur, Belgium \\
             \email{qkral@ast.cam.ac.uk, jean.schneider@obspm.fr}}

   \date{Received ; accepted }

  \abstract{Astrometry covers a parameter space that cannot be reached by RV or transit methods to detect terrestrial planets on wide orbits. In addition, high accuracy astrometric measurements are necessary 
to measure the inclination of the planet's orbits.  Here we investigate the principles of an artefact of the astrometric approach. Namely, the displacement of the photo-centre due to inhomogeneities 
in a dust disc around the parent star. Indeed, theory and observations show that circumstellar discs can present strong asymmetries.  We model the pseudo-astrometric signal caused by these inhomogeneities, asking whether a dust clump in a disc can mimic 
the astrometric signal of an Earth-like planet. We show that these inhomogeneities cannot be neglected when using astrometry to find terrestrial planets. 
We provide the parameter space for which these inhomogeneities can affect the astrometric signals but still not be detected by mid-IR observations. We find that a small cross section of dust corresponding to a cometary mass object is enough to mimic 
the astrometric signal of an Earth-like planet. Astrometric observations of protoplanetary discs to search for planets can also be affected by the presence of inhomogeneities. Some further tests are given to confirm whether an observation is a real planet astrometric signal or an impostor. Eventually, we also study the case where the cross section of dust is high
enough to provide a detectable IR-excess and to have a measurable photometric displacement by actual instruments such as Gaia, IRAC or GRAVITY. We suggest a new method, which consists in using astrometry to quantify asymmetries (clumpiness) in inner debris discs that cannot be otherwise resolved.}

      
\authorrunning{Kral, Schneider et al.}
\titlerunning{Effect of discs on astrometric measurements}

   \maketitle
%
%___________________________________________________________________

\section{Introduction}
The search for and observation of exoplanets is growing in three directions: new detection techniques are still emerging, ever more planets being discovered and more data being collected on each planet. The detection techniques have each their specificities regarding the accessible planet observables. 
Among them, astrometry is still in its infancy since, as of January 2016, only a handful of astrometric detections of exoplanet candidates have been announced, 
namely DE0823-49 b \citep{2015A&A...579A..61S} and  HD 176051 b \citep{2010AJ....140.1657M}. Another candidate, VB 10 b, claimed to be detected by astrometry \citep{2009ApJ...700..623P},
has been challenged by \citet{2010ApJ...711L..19B} and by \citet{2010ApJ...711L..24A}. We shall present a plausible reason that reconciles the three studies in the discussion.
This technique has yet a promising capability, which is to detect low mass planets, determine the geometry of their orbits,  and more importantly measure their masses. %It is the only method capable to detect and measure the mass of Earths/Super-Earths on wide orbits around nearby stars.

Just like any technique, astrometry has its own source of noise and artefacts. Until now, the only noise source limiting the sensitivity to low masses that was investigated is the activity of the central star. The latter introduces indeed a fluctuation of the centroid of the star position \citep{2011A&A...528L...9L}.

Here we investigate an artefact of the astrometric approach, first emphasised by \citet{Schneider2011}, whose preliminary results showed that an astrometric signal subject to a perturbation by an axially asymmetric dust disc can mimic the dynamical effect of planetary companions.

This study is important in the context of astrometric projects aiming at the detection of Earth-like planets in the Habitable Zone of nearby stars \citep[e.g. NEAT,][]{2012ExA....34..385M}. On the ground, the GRAVITY \citep{2014SPIE.9146E..2EL} and SKA \citep{2004NewAR..48.1473F} instruments or projects
have an astrometric accuracy of a few microarcsec. For Earth-like planets, the astrometric signal expected is on the order of 0.3 $\mu$as around a sun-like star at 10 pc (see Eq.~\ref{eqB}) and artefacts of this order of magnitude could mimic such planets. Therefore they have to be taken into account.

We first present the general mathematical formalism in section \ref{impa} and show how it works on a range of cases in section \ref{spech}. In section \ref{discu} we provide some tests to discriminate a disc-induced artefact from a real stellar wobble. Finally,
we compare the effect of inhomogeneities to the performance of present and future astrometric instruments such as Gaia \citep{2012Ap&SS.341...31D}, GRAVITY, TMT \citep{2013JApA...34...81S} and NEAT. Furthermore, we discuss a new interesting detection technique to observe clumpiness in discs.

\section{Impact of inhomogeneities on parent star astrometry}\label{impa}

In this section we consider the case of stellar light reflected by some dust inhomogeneities as well as its thermal emission. The point of using astrometry is to reach accuracy below the resolution of the system. In this study, we are thus interested in inhomogeneities that cannot be
resolved. For this reason, the dust creating these inhomogeneities must be close to its host star. For an observation at a wavelength $\lambda$ with a telescope of diameter $d$, the typical spatial resolution is $\lambda/d$ meaning that for NEAT in the optical, the resolution is $\sim$ 0.1 arcsec.
As a result, inhomogeneties that would affect astrometric measurements without being resolved will be within 0.1'', i.e. within 1au (10au) for a system at 10pc (100pc).

The brightness distribution of inhomogeneities is a function of time since the disc structure evolves essentially according to its Keplerian motion. Thus, the difference in position between the star and the displacement due to the presence of an asymmetric disc can be given as a function of 
the inhomogeneity to star flux ratio $I_{\rm inh}/I_{\star}$. If a companion is present, it also changes the observed position of the star and this is 
what is used to detect planets with astrometric measurements.

\begin{figure}
   \centering
   \includegraphics[width=8.5cm]{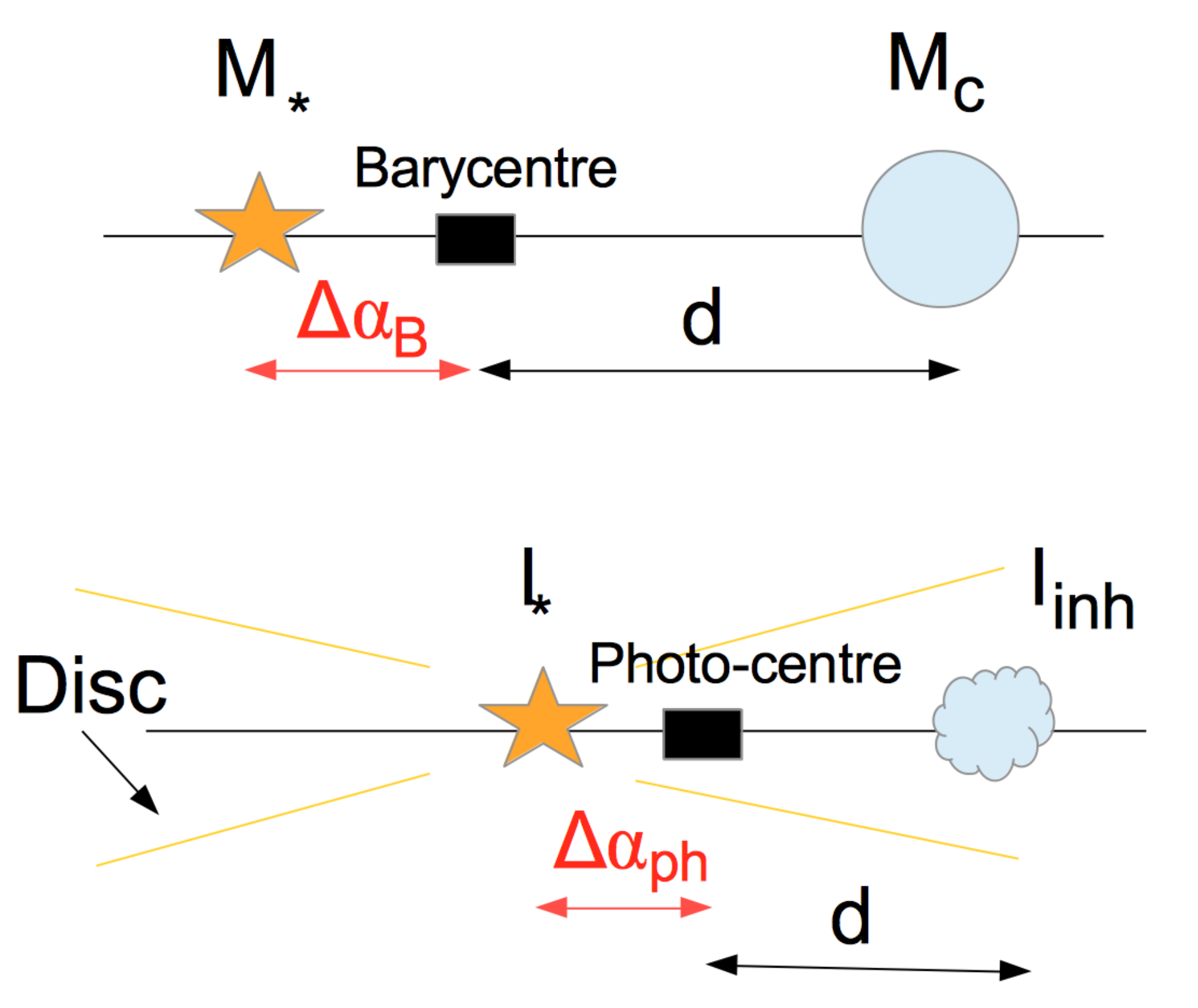}
   \caption{\label{schema} {\it Top:} Schematic of the usual barycentric photo-centre displacement $\Delta \alpha_B$ used in astrometry to detect a planet of mass $M_C$ at distance $a$ from the barycentre with the star of mass $M_\star$. 
{\it Bottom:} The photo-centre displacement $\Delta \alpha_{ph}$ is now caused by an inhomogeneity of flux $I_{\rm inh}$ at distance $a$ orbiting around a star of flux $I_\star$ and possibly surrounded by a disc shown with yellow lines.}
\end{figure}

As astrometry is performed below the resolution limit, we observe the displacement of the photo-centre of the system. The global photo-centre displacement $\Delta \alpha$ is a superposition of the barycentric dynamical astrometric displacement  $\Delta \alpha _B$ and the photo-centric displacement $\Delta \alpha _{ph}$ due to the asymmetric brightness distribution of the 
unresolved disc (see Fig.~\ref{schema}).

Let $\alpha_{ph}$ be the angular separation of the photo-centre of an inhomogeneity in the disc, of brightness ratio $I_{\rm inh}/I_{\star}$,  with respect to the parent star. Then the star$+$inhomogeneity photo-centre position has an angular offset $\Delta \alpha_{ph}$ with respect to the parent star's centroid given by

\begin{equation}
\label{eqalpha}
\Delta \alpha _{ph}=  \frac{I_{\rm inh}}{I_{\star}} \alpha_{ph}.
\end{equation}

By definition, neglecting the mass of the disc inhomogeneities, the barycentric displacement due to a companion with mass $M_C$ gives

\begin{equation}
\label{eqB}
\Delta \alpha _B = \frac{M_C}{M_\star} \alpha_{C},
\end{equation}

\noindent where $\alpha_{C}=a/D$ is the angular separation of the companion, $a$ is the barycentre-planet distance projected on the sky at the measurement epoch and $D$ the distance of the star to the observer.

Let us now assume that the photo-centre displacement is created by an inhomogeneity at distance $a$ so that $\alpha_{ph}=a/D$ and the amplitude of the displacement $\Delta \alpha =  I_{\rm inh}/I_{\star} (a/D)$, where $a$ is in au, $D$ in pc and $\Delta \alpha$ in arcsec.

\begin{figure*}
   \begin{minipage}[c]{90mm}
      \includegraphics[width=8.9cm]{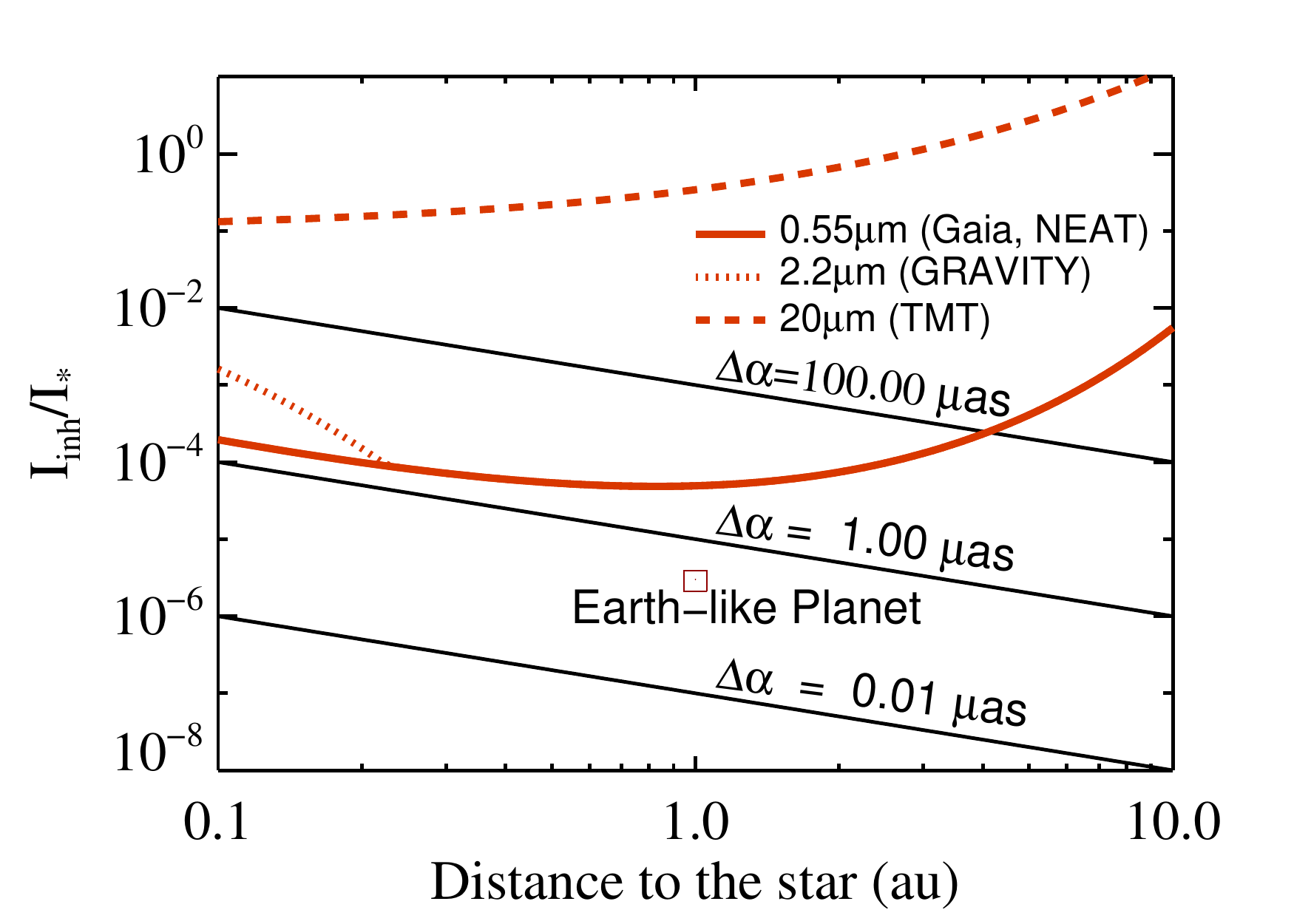}
  % \caption{\label{figA} Simulation A \textbf{(image to be redone in cgs and right injection rate)}: Surface density profile at different epochs. $t=[5 \times 10^4, 10^6, 10^8, 10^9]$ years.}
   \end{minipage} \hfill
   \begin{minipage}[c]{90mm}
      \includegraphics[width=8.9cm]{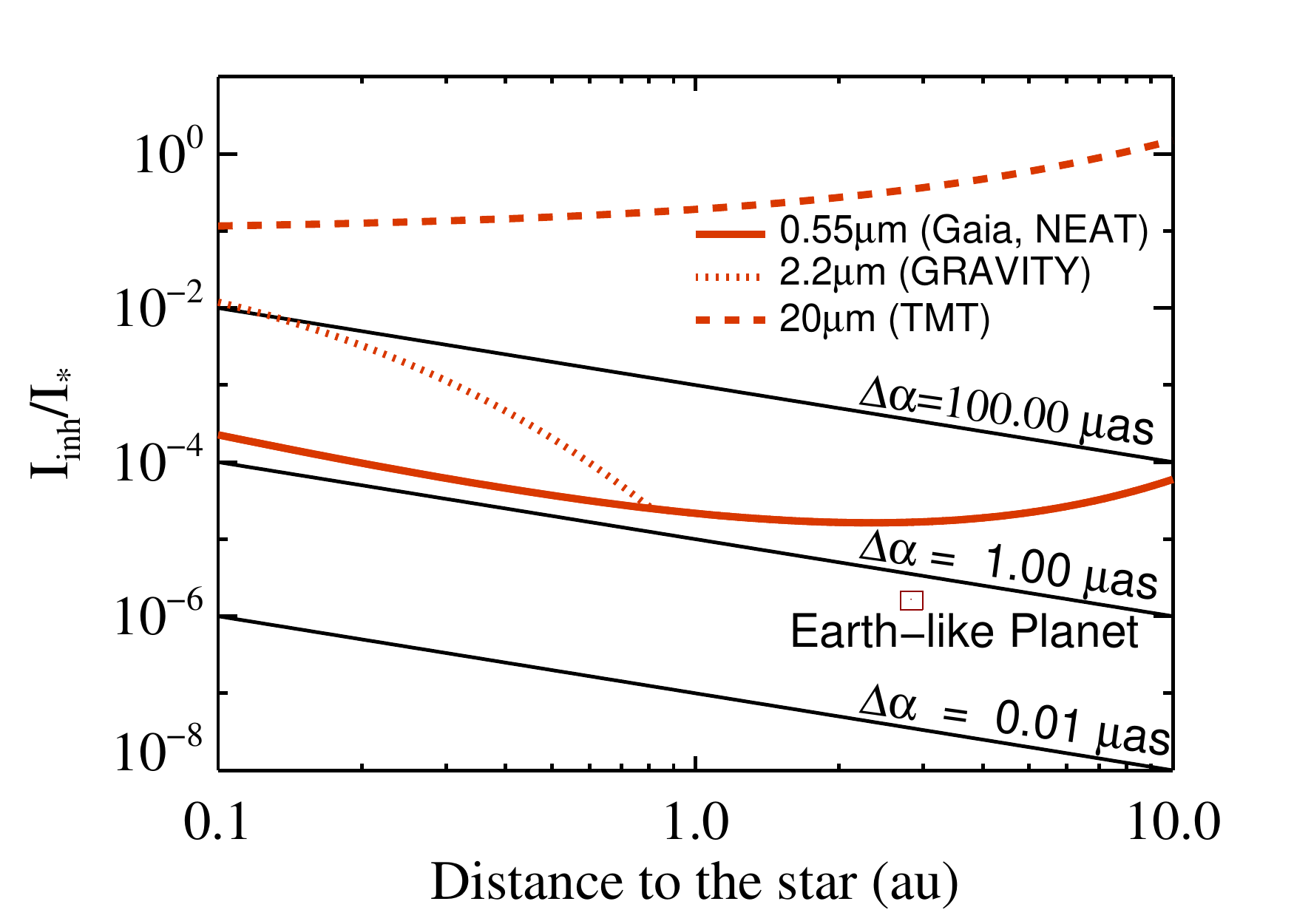}
     \end{minipage}
   \caption{\label{figA} {\it Left:} $I_{\textrm{inh}}/I_{\star}$ as a function of the distance to the star $a$ (in au) for different photo-centre displacements ($\Delta \alpha=[0.01,1,100]$ $\mu$as) in black solid lines. WISE (12$\mu$m) detection limit in terms of $I_{\textrm{inh}}/I_{\star}$
 are shown in red for a dust cloud orbiting at $a$ around a sun-like star and observed at different wavelengths representing different potential instruments, 0.55 (solid, Gaia or NEAT), 2.2 (dotted, GRAVITY) and 20 (dashed, TMT) microns. The position of an Earth-like planet is represented with a red square.
{\it Right:} Same for a dust cloud orbiting a $\beta$-Pic like A6V star. In this case, the Earth-like planet having the same irradiation and mass as Earth is located at 2.83au (see text for details).}
\end{figure*}

First, we plot on Fig.~\ref{figA} $I_{\rm inh}/I_{\star}$ as a function of $a$ assuming a distance $D=10$ pc, for different values of $\Delta \alpha$ (black solid lines). As given by Eq.~\ref{eqalpha}, when $I_{\rm inh}/I_{\star}$ increases, for a given $a/D$, $\Delta \alpha$ increases. Also, for a given $\Delta \alpha$, 
$I_{\rm inh}/I_{\star} \propto 1/a$. The location of an Earth-like planet in terms of its $\Delta \alpha$ needed at 1au to mimic an Earth is shown as a red square in Fig.~\ref{figA}. This gives the inhomogeneity-to-star flux ratio required to mimic an Earth. On the right plot, this location changes as
the host star is more massive by a factor $\sim$ 1.8. Indeed, using Eq.~\ref{eqB}, we can rewrite it in these more useful terms:

\begin{equation}
\label{eqas}
\Delta \alpha _B = 0.3\mu\textrm{as} \, \frac{M_p}{M_\oplus} \left( \frac{M_\star}{M_\sun} \right)^{-1} \frac{a'}{1 {\rm au}} \left( \frac{D}{10 {\rm pc}} \right)^{-1}.
\end{equation}

\noindent Simply assuming that habitable planets around more or less luminous stars reside at distances where the insolation is equal to the Earth’s, then $a'=a \, (L_\star/L_\sun)^{0.5}$ is the new location of the planet. The location of an Earth-like planet around $\beta$ Pic is then at $a'=2.83$ au and would imply a displacement of 0.47$\mu$as.

We assess the detectability of such inhomogeneities
using the WISE detection limits at $\lambda_{\rm lim}$=12$\mu$m. At a given astrometric observing wavelength $\lambda_{\rm inh}$ we compute the minimum value of $I_{\rm {inh}}/I_{\star}$ that could be detected by WISE at different $a$. 
To do so, we assume that the detections are limited by our ability to differentiate the WISE photometry from the stellar flux (``calibration limited'') and that the grains behave like pseudo-black bodies
with an albedo $\omega=0.2$. We find that in thermal emission, the minimum detected flux ratio at wavelength $\lambda_{\rm inh}$ is given by

\begin{equation}
\frac{I_{\rm inh}}{I_\star}(\lambda_{\rm inh})=R_{\rm lim} \frac{B_\nu(\lambda_{\rm inh},T)}{B_\nu(\lambda_{\rm inh},T_\star)} \frac{B_\nu(\lambda_{\rm lim},T_\star)}{B_\nu(\lambda_{\rm lim},T)},
\end{equation}

\noindent where $B_\nu$ is the Planck function and $R_{\rm lim}$ is the WISE calibration limit ($\sim 10\%$). To get this formula, we compute the disc-to-star flux ratio at a certain wavelength $\lambda_{\rm lim}$ set by the WISE sensitivity and shift it to another
(thermal) wavelength $\lambda_{\rm inh}$ where the astrometry is done. The albedo does not appear in this equation as it affects both the WISE and the astrometric wavelengths equally. Similarly, in scattered light, converting the thermal detection limit to an optical flux, this limit is defined by

\begin{equation}
\frac{I_{\rm inh}}{I_\star}(\lambda_{\rm inh})=6\times 10^9 \frac{\omega}{1-\omega} \frac{R_{\rm lim}}{a^2} \frac{B_\nu(\lambda_{\rm lim},T_\star)}{B_\nu(\lambda_{\rm lim},T)} \frac{L_\star}{T_\star^4},
\end{equation}

\noindent where $L_\star$ is the star's luminosity in solar luminosity, $T_\star$ the star's temperature (in K), and $a$ is in au. $T=278.3 \, L_\star^{0.25}/\sqrt{a}$ is the black body temperature of the inhomogeneity. We assumed that the absorption/scattering properties are wavelength independent (i.e. albedo is monochromatic)
but we illustrate what happens in Fig.~\ref{deltacloud} for a full Mie calculation.

The red curves on Fig.~\ref{figA} show the corresponding limits for different observing wavelengths $\lambda_{\rm inh}$ equal to 0.55 (solid line, for Gaia, NEAT), 2.2 (dotted line, for GRAVITY) and 20 (dashed line, for TMT) $\mu$m. The left plot is for the case of a sun-like star and the right plot for a $\beta$ Pic-like
star. In this last case, we consider an Earth-like planet as having the same irradiation and mass as Earth, so it is located at 2.83au as stated earlier when defining $a'$. Above each red line, the dust inhomogeneity could be detected by WISE at 12$\mu$m. One can notice the wide range of the parameter space where some dust could be undetected but still create a photo-centre displacement big enough to mimic an Earth-like planet. The limits on the amount of dust depend on the type of host star (which changes the dust temperature)
as well as $\lambda_{\rm inh}$. For observations of the astrometric displacement in the optical, the emission is dominated by scattered light whilst at 20$\mu$m, only thermal emission is present. For the $\lambda_{\rm inh}$ = 2.2 microns case shown on the plot, both thermal emission and scattered light can be important
depending on the distance to the star, which explains the two regimes observed on the dotted line. Indeed, for a sun-like ($\beta$ Pic like) star, within 0.2au (0.8au), the grains are so hot that thermal emission becomes dominant over scattered light. We conclude here that there is still a large parameter space (which gets larger for lower mass stars) where inhomogeneities can be bright enough to induce a photo-centre displacement but still be
undetectable. Of course, the detection of an IR-excess in a system with an astrometric detection does not guarantee that the dust is the cause, as the dust distribution could be axisymmetric.

\section{Various cases of inhomogeneities in planetary systems}\label{spech}
\subsection{Overview of the different cases}\label{over}

The inhomogeneities presented in the previous section can be classified in two different categories of axial asymmetries in discs: static and non static. Beyond that general dichotomy, it is not trivial to describe the diversity
of asymmetric morphologies by a limited number of categories. One could imagine diverse situations such as ``bright spots'', ``vortices'', ``spiral structures'' or even ``giant impacts''. The static inhomogeneities lead to a photo-centre displacement that is constant in time and can therefore not mimic the stellar motion due to a companion.
Thus, the only inhomogeneities that we are interested in are non static configurations. For this reason, we discard giant impacts, which show a strong asymmetry at the collision point at steady state but that does not move and could be distinguished from a planet \citep{2015A&A...573A..39K}. 
For the same reasons, the warps in discs (created by an inclined planet) and pericentre glow effects (brightness asymmetry due to secular perturbations by an eccentric planet, leading to an eccentric disc) are discarded. The interesting cases that could affect astrometric measurements can
be summed up as follow:

\begin{itemize}
 \item 1) Bright spots (dust cloud in a disc, clumps due to a resonance with a planet),
 \item 2) Spiral structures (due to a planetary companion or a fly-by with a nearby star),
 \item 3) Inhomogeneities in protoplanetary discs (vortices, planet induced spirals, ...)
 \item 4) Non static others
\end{itemize}

Let us now parametrise the different non static configurations listed above relaxing the previous black body assumption and assess their respective impacts on astrometric observations.

\begin{itemize}
\item ``Bright spots'' \\

Bright spots are observed in discs and are expected from theory. These bright spots could be dust clouds that are created by dust produced in collisions between planetesimals within an otherwise undetectable debris disc \citep{2002MNRAS.334..589W,2014MNRAS.439..488Z}.
One could also imagine that the spots are dust clumps that stay within the Hill sphere of planetesimals. These enshrouded planetesimals are surrounded by a swarm of irregular satellites that collide and create the dust cloud \citep{2011MNRAS.412.2137K}.  
Another possibility would
be that planetesimals are surrounded by rings such as the centaur Chariklo \citep{2014Natur.508...72B}, but more massive and with a high albedo. These clumps are also expected trailing behind planets, as in the Earth's resonant ring that corotates with Earth \citep{1999ASPC..177..374W}. More generally, dust grains can be trapped in mean-motion resonances with planets when migrating inwards due to Poynting-Robertson drag, which can create luminous dust clumps \citep[e.g.][]{2015MNRAS.448..684S}.
We do not intend to be exhaustive here but rather give an idea of several configurations that could create these bright spots.

In this case, we can use Eq.~\ref{eqalpha}, where $I_{\rm inh}=I_{spot}(a_s,\lambda)$, the spot flux at wavelength $\lambda$ and at a projected distance $a_s$ that we shall assume to be the semi-major axis of the dust cloud for simplicity. 
Also, $\alpha_{ph}=a_s/D$, where $D$ is the distance to Earth. $I_{spot}$ is the sum of thermal emission $I_{spot_{th}}$ and scattered light $I_{spot_{sca}}$. $I_{spot_{th}}$ (assuming a fixed composition) is equal to

%\begin{multline}
%\label{Is}
 %I_{spot_{th}}(s,a_s,\lambda) =\int_{s_\textrm{min}}^{s_\textrm{max}}  \frac{\sigma_\textrm{abs}(\lambda,s)}{4 D^2} \\
% B_\nu(T_s(a_s,s)) \, \textrm{d}n(s),
%\end{multline}

\begin{equation}
\label{Is}
I_{spot_{th}} =\int_{s_\textrm{min}}^{s_\textrm{max}}  \frac{\sigma_\textrm{abs}(\lambda,s)}{4 D^2}  B_\nu(T_s(a_s,s)) \, \textrm{d}N(s),
\end{equation}

\noindent where $\sigma_\textrm{abs}$ is the absorption/emission cross section of the cloud for a given grain size $s$, i.e $4 \pi s^2 Q_{\rm abs}(s,\lambda)$ and $B_\nu(T_s)$ is the Planck function for a temperature $T_s$.  
$Q_{\rm abs}$ is the dimensionless absorption/emission coefficient, which for a given grain size depends only on the wavelength $\lambda$. $\textrm{d}N(s) \propto s^{-q} \, \textrm{d}s$ is the size distribution of grains. We choose the standard $q=3.5$
value to model dust clouds \citep[e.g.][]{2013A&A...558A.121K}. The grain temperature $T_s$ is worked out
solving the thermal equilibrium of grains using the code GRaTeR \citep{1999A&A...348..557A,Lebreton2013}. Indeed, for small grains the black body assumption is far from being accurate and as we are interested in the whole wavelength range for astrometric measurements, 
we assess the infrared emission from the dust properly through Eq.~\ref{Is}.

Similarly, for the case of scattered light
\begin{equation}
\label{Iss}
I_{spot_{sca}}(a_s,\lambda) = I_{*}(\lambda) f(\phi) \frac{\sigma_\textrm{sca}(\lambda)}{a_s^2},
\end{equation}

\noindent where $\sigma_\textrm{sca}(\lambda)=\int_{s_\textrm{min}}^{s_\textrm{max}} \pi s^2 Q_\mathrm{sca}(s,\lambda) \, \textrm{d}N(s)$, $Q_\mathrm{sca}$ being the dimensionless scattering coefficient and $f(\phi)$ is the phase function, which is equal to $1/4\pi$ as we assume isotropic scattering.\\
 
\item ``Spiral structures'' \\

Two types of spiral structures can develop in debris discs. Spirals due to tidal forces following the passage of, for instance, an unbound companion or over longer timescales due to differential precession for the case of a bound companion. The tidally induced
spiral evolves quickly as its evolution is set up by the differential rotation of the disc $\textrm{d}\Omega/\textrm{d}R$, where $\Omega$ is the orbital frequency and $R$ the distance of planetesimals. The probability to witness such spirals is thus 
rather low and we do not intend to model them. For the other type of spiral, the evolution is set by the differential precession 
of the orbits $\textrm{d}\omega_p/\textrm{d}R$, where $\omega_p$ is the precession velocity of the orbits. One finds that for a circular perturber of mass $M_p$ and semi-major axis $a_p$ \citep{2004A&A...414.1153A}

\begin{equation}
\label{omegap}
\omega_p=\frac{3 G M_p}{4 \Omega a_p^3}.
\end{equation}

The evolution of this type of spirals is then very slow compared to the Keplerian evolution and could not be misinterpreted as being a planet.\\

\item ``Inhomogeneties in protoplanetary discs''\\

We created a general category for these young systems still surrounded by a protoplanetary disc and we give some specific cases that could interfere with astrometric measurements.

Vortices are expected to be present in protoplanetary discs being induced either by the Rossby-wave instabilities \citep[RWI,][]{1999ApJ...513..805L} or baroclinic instabilities \citep{2010A&A...513A..60L}. 
In the case of the RWI, it can be triggered either on the edge of a gap carved by giant  planets or at the interface with dead zones. These vortices can even be created very close to the star at the inner disc edge at the boundaries
between turbulent, magnetised and accreting regions \citep{2015A&A...573A.132F,2016ApJ...816...69A}. These vortices can survive for several orbits which can be shortened depending on several 
complicated effects such as, for instance, the amount of turbulence or dust feedback effects \citep{2014ApJ...788L..41F}. 
One such vortex was plausibly observed by 
ALMA \citep{2013Sci...340.1199V}. These vortices can be, as a first approximation, modelled as we have done for dust clouds. As large dust concentrates in the vortex centre, we shall take a shallower size distribution within the affected size range 
as explained in \citet{2013ApJ...775...17L}.

%We create a 1D model of vortices as we are not interested in the spatial distribution of grains. However, when the Stokes number (St $=\tau_s \Omega=\sqrt{\pi/8}s \rho/(H \rho_g)$) is close to 1 ($\tau_s$ being the stopping time, $\rho$ the bulk density of grains and $\rho_g$ the gas number density),
%the dust shows an overdensity, which might affect our results if the corresponding grain size $s$ is small enough to affect images. To be more accurate, \citet{2013ApJ...775...17L} find that the maximum density evolves $\propto (100 \, \textrm{St}+1)^{1.5}$
%(we assumed that the turbulent velocity is 10\% of the sound speed in their Eq.~63). This density evolution can be translated in a size distribution where $\textrm{d}N \propto s^{-3} \textrm{d}s$. We fix our maximum grain size $s_{\rm max}$ where St=1. 
%We place our cut-off at St=0.01 where we stop using the Eq.~63 mentioned above but instead take a size distribution in $s^{-3.5}$. Indeed, when the grains are smaller, they end up being totally coupled with the gas and we expect an ISM-like
%size distribution. We fix the minimum grain size to be 1 micron. We normalise the whole distribution by a vortex dust mass $M_v$ composed of grains up to $s_{\rm max}$.\\

Planet induced spirals \citep{2002ApJ...565.1257T} are expected in protoplanetary discs and one could end up having both the contribution of the planet and the spirals in the astrometric measurements. Gravitational instabilities may develop in protoplanetary discs and lead to fragmentation and
clump formation \citep[e.g.][]{2015MNRAS.454.2529M}. SAO 206462 is an example of young disc presenting spiral features that could be generated with either a planet or gravitational instability \citep{2013A&A...560A.105G}. 
Spiral waves may also be triggered by shadows in transition discs \citep{2016arXiv160107912M} or even by an inflow coming from the residual external envelope of protoplanetary discs \citep{2015A&A...582L...9L}.

We also note that mid-IR variability has been observed in many protoplanetary discs \citep{2014ApJ...793....2F}. This may be due to variations in the structure of the inner rim of the protoplanetary disc. Depending on the disc orientation, this would also cause an astrometric effect.
The astrometric displacement derived from such a variation would also depend on the inclination, and might not look like an ellipse however, since the bright spot (the puffed up bit of rim) would not necessarily be visible for the whole orbit.\\

\item ``Others''\\

This subsection includes the case of many different substructures that can be found in planetary systems that shall be treated on a case by case basis. For example, the case of AU Mic shows that we do not yet fully 
understand all types of inhomogeneities \citep{2015Natur.526..230B}. In this system, some dust clumps are observed to move radially outwards with time. It can be modelled as a dust clump that is composed of unbound grains and move radially outwards.\\

\end{itemize}
   
\subsection{Quantify the effect of inhomogeneities for some specific cases of debris and protoplanetary discs}\label{observed}
We now evaluate the quantitative effect of disc asymmetries listed in subsection \ref{over}. Since we have in mind astrometric programs with timescales less than 10 years, we must consider the inner part of discs closer than 1 arcsec of the parent star. 
Such regions have not yet been investigated, with the exception of a few interferometric observations \citep[e.g. Fomalhaut,][]{Lebreton2013} and photometric light curves \citep{Rodriguez2015}. 
We are thus constrained to rely on models extrapolating below 1 arcsec (assuming systems close to the Earth) observations made further than 1 arcsec of the parent star.

\begin{itemize}
\item ``Bright spots''\\

For this example, we shall model a dust cloud orbiting at 1au around a sun-like star at 10pc. We would like to check that Fig.~\ref{figA} still holds when relaxing the black body assumption. A signal to be mistaken with an Earth-like planet, with the same orbital period and the same astrometric 
amplitude would create a 0.3$\mu$as astrometric signal. It converts in 
$I_\textrm{inh}/I_\star=3 \times 10^{-6}$ or a total cross section $\sigma_{\rm tot} \sim 10^{18}$ m$^2$. Thus, we input a dust cloud with a cross section of $10^{18}$ m$^2$. 
The minimum grain size in the dust cloud
is set by radiation pressure cut-off, which is $\sim$ 1 micron for this type of host star. We fix the maximum size as being 100m, but one should note that the biggest bodies will not contribute to the flux at all. This fixes the total mass to be $3 \times 10^{19}$ kg (30 times
lighter than Ceres). We use a 
conventional size distribution with $q=3.5$. The photometric displacement due to the dust cloud depends on the wavelength as shown in Fig.~\ref{deltacloud}. It does create a $\Delta \alpha=0.4\mu$as in the optical, which is close to the expected 0.3$\mu$as in the optical. 
For longer wavelength, the astrometric displacement is even larger. We overplot the results of our analytical model for varying spectral types (A6V, G2V and M8V) in Fig.~\ref{deltacloud}. The values predicted with our black body assumption provide a very good estimate of the real displacement. We note
two major differences. First, by comparing the numerical model (thick line) and the analytical black body assumption, one notices that they behave differently at longer wavelengths because grains are inefficient emitters when $\lambda > s$. Secondly, the break between the scattered light
dominated regime (smaller $\lambda$) and the thermal emission dominated regime happens at longer $\lambda$ for latter spectral types (see subsection \ref{chroma} for more details).

\begin{figure}
   \centering
   \includegraphics[width=8.5cm]{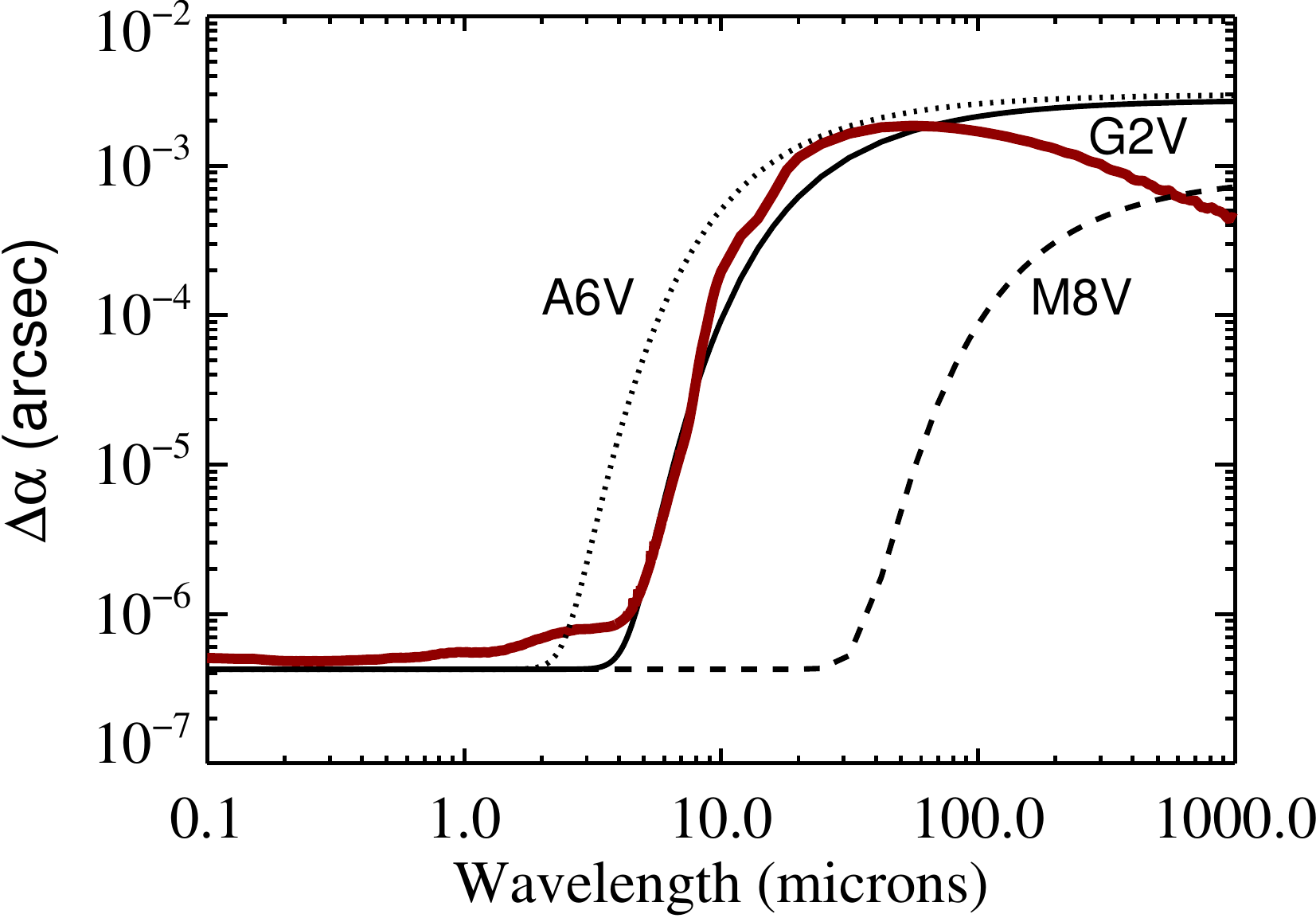}
   \caption{\label{deltacloud} Photo-centre displacement as a function of wavelength for a Keplerian dust cloud orbiting at 1au around a G2V star at 10pc (solid lines). The thicker of the G2V lines is the numerical model. We also vary the spectral type of the host star using 
the analytical model, the dotted line is for an A6V star and the dashed line for an M8V star.}
\end{figure}

This implies that an undetectable cometary mass body, if broken up and distributed as a collisional cascade size distribution, is enough to create a signal greater than 0.3 $\mu$as (which is the astrometric displacement due to an Earth-like planet at 1au at 10pc) in the optical. This dust cloud
is then able to mimic an Earth-like planet as would be any more massive dust clouds. The mass is too small to change the barycentre position significantly so that the only contribution to the photo-centre displacement is from the dust cloud brightness.\\

\item ``Inhomogeneities in protoplanetary discs''\\

Most of the inhomogeneities presented in subsection \ref{over} for protoplanetary discs could be modelled with a bright spot.
For instance, vortices could easily be modelled with a dust cloud with two different size distributions as the biggest grains, with a Stokes number close to one, are most easily trapped.
We emphasise however that these features are surrounded by a protoplanetary discs that should definitely be detected if present (e.g. by IR-excess). 
 Anyway, all the cases presented in subsection \ref{over} might arise, but since most of them can be modelled with a bright spot, a more refined study is left for the future when astrometric data of systems hosting a protoplanetary discs will be available.

%An ALMA observation of the system Oph IRS 48 \citep{2013Sci...340.1199V} could be explained with the presence of a vortex dust trap.
%We use the parameters derived from this observation to fix our free parameters. The peak density of the ALMA observation is at about 63AU but if we assume that the gas density evolves $\propto R^{-1}$ and the scale height as $R$ so that the Stokes number
%does not vary with distance to the star and their derived most efficient trapping size can be applied for a vortex that would be at a few AU. From the observation, we set $s_{\rm max} \sim 5$mm, which gives the cut-off size at St=0.01 
%equal to 50 microns and we set the size distribution slope to $q=3$. Note that $s_{\rm max}$ depends on the parameters chosen for the disc \citep[e.g. 0.3mm in][for Oph IRS 48]{2014ApJ...795...53Z}. $M_v$ $\sim$ 9 M$_\oplus$ from the ALMA observation but we keep it as a free parameter and set a minimal mass that can mimic the signal of an Earth-like planet. 

We use Fig.~\ref{figA} (right) to see that one needs $I_{\rm inh}/I_\star$ to be $\sim$ 1.5$\times 10^{-6}$ to mimic an Earth-like planet around an A6V star ($a'=2.83$au) at 10pc, which translates as a total cross section equal to $2\times 10^{18}$ m$^2$. Using this cross section, we check with GRaTeR
 that the results are very similar to those presented on Fig.~\ref{deltacloud}. Such a cross section is sufficient to imply a displacement of the photo-centre large enough to perturb the detection of an Earth-like planet. This confirms once again that our black body approximation on Fig.~\ref{figA} gives sensible results. Indeed, the photo-centre 
displacement created by the inhomogeneity emission is equal to the expected 0.45$\mu$as (see Eq.~\ref{eqas}) in the optical and is greater at longer wavelength as is found for the bright spot case (see Fig.~\ref{deltacloud}). 
The displacement induced at longer wavelengths could be detected by high accuracy astrometric instruments such as SKA or even using IRAC/Spitzer that can reach an astrometric precision of $\sim$ 20mas \citep{2016AJ....151....9E}. 
This shows that astrometric observations of systems hosting protoplanetary discs may be affected by these asymmetric structures. Disentangling a real planet from these inhomogeneties requires extra work as explained in details in section \ref{discu}. \\

%\begin{figure}
%   \centering
%   \includegraphics[width=8.5cm]{deltaalpha2.jpeg}
%   \caption{\label{deltavortex} Photo-centre displacement as a function of wavelength induced by a Keplerian vortex orbiting at 2.83AU around an A6V star at 10pc.}
%\end{figure}

\item ``Others''\\

As an example of ``other'' configurations we take the disc around AU Mic. It shows fast, apparently non Keplerian, motions of spots.  Assuming a contrast ratio of $10^{-5}$ relative to the parent star corresponding to the high contrast 
imaging performances of the images of AU Mic with SPHERE \citep[see][]{2015Natur.526..230B}, the features A, B, C, D and E lead respectively to the following pseudo astrometric effects in $\mu$as (derived from their Figure 3b) at epochs 2011 (HST) and 2014 (SPHERE):\\

\begin{tabular}{c|c|c|c}
Image & $\alpha_{ph}$ 2011 & $\alpha_{ph}$ 2014 & $\Delta \alpha_{ph}$ 2014-2011\\ \hline
A & 8 &  10 & 2 \\
B & 35 &  42  & 7 \\
C & 60 &  68 & 8 \\
D &  72 &  76 & 4 \\
E &  90 &  96 & 6 \\
\end{tabular}\\

Assuming an astrometric instrument with a 1 arcsec resolution, only the spine A would have an astrometric effect on the parent star. It would lead an astrometric difference between 2011 and 2014 of $2 \mu$as. It corresponds to an astrometric effect linear in time between 2011 to 2014;  it can therefore not mimic the Keplerian motion of a companion with a period of only a few years, since the transverse velocity of the latter would be sinusoidal on an edge-on orbit. 
\end{itemize}

\section{Tests to confirm an astrometric planet detection}\label{discu}
Here we discuss some observational tests to check if an apparent photo-centre displacement is due to a barycentric effect by a companion or to a disc asymmetry.

\subsection{Comparison with radial velocity measurements}

Some bright non static asymmetries could mimic a pseudo-planet sufficiently massive to be detectable by radial velocity measurements. A first trivial check would therefore consist
in a monitoring of the radial velocity of the parent star, if permitted by the stellar spectrum.

The proposed test has been applied by \citet{2010ApJ...711L..24A} and by \citet{2010ApJ...711L..19B} to the planet
candidate claimed by \citet{2009ApJ...700..623P} to be detected by astrometry. These authors did not detect it by radial velocity, concluding that VB 10 b does not exist. We suggest that a possible explanation of the contradiction
 between the astrometric claim by Pravdo \& Shaklan and the negative result from radial velocities measurement consists in an astrometric effect due to an asymmetric disc inhomogeneity. We recomputed Fig.~\ref{figA} for this case, where the host star
has very low luminosity (M8V star) and temperature ($\sim$ 2700 K). The detection limit given by WISE for this case is very high, which would allow a massive dust cloud creating a $\Delta \alpha_{ph}$ greater than 1mas to be present without being detectable. \\

\subsection{Time evolution}

If the photo-centre evolution is a stellar wobble due to a companion, it must correspond to a Keplerian motion. Such Keplerian motions are also the case for bright spots and vortices, while the cases of AU Mic or spiral evolution show that for some cases the effective photo-centre 
displacement may be non-Keplerian. A first test is thus to check if the displacement is Keplerian. If it is not, the wobble is not due to a companion.

Nevertheless, it is not as straightforward if there are more than one planetary companion with different periods, since the stellar wobble is then different from a simple Keplerian motion. Vice versa, if the displacement is Keplerian, 
it is not a proof that it is a dynamical stellar wobble as explained in this paper. The eccentricity of the object creating the astrometric signal may give some clues as for the favoured scenario: planet versus dust. Indeed, one may expect dust clouds to have 
a rather low eccentricity but planets can reach higher values. This could not be claimed as a secured test but give some hints to disentangle the different possibilities.\\

\subsection{Chromaticity}\label{chroma}

The dynamical stellar wobble induced by one (or more) companion is wavelength independent. In particular it must be the same in reflected light (in the visible and near-infrared domain)
and in thermal regime (mid-infrared to mm). An achromaticity\footnote{We here define a chromatic signal as being wavelength dependent and an achromatic signal as being constant with varying wavelength} of the photo-centre displacement is a strong test of its purely dynamical origin. On the contrary, an inhomogeneity would create a chromatic signal (e.g. see Fig.~\ref{deltacloud}). Therefore, whenever possible, a candidate astrometric displacement should be observed in at least 
two wavelength regimes where a change is expected. The two observing wavelength must be chosen accordingly to the spectral type of the host star as the break between a reflected light dominated regime and a thermal emission regime happens at longer wavelength for latter spectral types (see Fig.~\ref{deltacloud}). For M dwarfs, one should look in the far-IR to mm to see a difference, whereas for an early A type star, NIR or mid-IR observations will
already show a difference. In any case, observing at longer wavelengths seems more favourable and SKA should be a good instrument to realise these observations. \\

In the above subsection looking at comparisons with radial velocity measurements we have suggested that the pseudo astrometric detection of VB 10 b is due to an asymmetric disc inhomogeneity. A confirmation of this explanation 
would be that the detection of the astrometric effect detected by Pravdo \& Shaklan at 550-750 nm is wavelength dependent, for instance, in the far-IR or submillimetric regimes. We recommend to observe it with, {\rm SKA, when built} using astrometry or the Large Binocular Telescope Interferometer 
\citep[LBTI,][]{2016arXiv160106866D} to look for warm dust.\\

\subsection{Direct imaging}

Direct imaging can be used a posteriori for a confirmation or rejection of the candidate astrometric signal. Indeed, a very strong test would consist in a high angular resolution image of the stellar environment to investigate possible non static asymmetries in a disc. 
Also, this is a good way to disentangle an astrometric signal coming from both an inhomogeneity and a planet (the achromoticity VS chromaticity of the signal could also be used), as would be created by an equivalent to the Earth's resonant ring that corotates with Earth \citep{1999ASPC..177..374W}.
In the visible and near-infrared the GPI camera on the Magellan telescope and the SPHERE camera on the VLT and in the millimetric regime the ALMA instruments are well adapted to this task. However, direct imaging of discs is only possible
for the brightest systems with a fractional luminosity greater than about $10^{-4}$ and there are not yet any cases where a disc has been imaged but no IR-excess was detected. As for protoplanetary discs, seeing structures within a few au is not yet possible for the distances
they reside at. Therefore, direct imaging could only be useful to detect the supposed planet otherwise detected by astrometry.

\section{Discussion}
Note that a byproduct of our discussion of the wavelength dependence of a pseudo-astrometric effect is to suggest to detect astrometrically inner discs not detectable by imaging and to constrain their asymmetry. 
It could in particular first be applied to some systems where Falling Evaporating Bodies (FEBs) are known like $\beta$ Pictoris or HD172555 \citep{1998A&A...338.1015B, 2014A&A...561L..10K}. It could also be tried on systems where
interferometric observations seem to show the presence of hot dust called exozodis \citep{2013A&A...555A.104A, 2015Msngr.159...24E}. The final aim would be to apply this astrometric method to randomly probe the inner parts of planetary systems
and discover some hidden components such as FEBs, asymmetric exozodis or even trailing clumps behind planets such as the Earth's resonant ring that corotates with Earth \citep{1999ASPC..177..374W}.

Hence, while we have shown that unseen dust in the inner regions of planetary systems may masquerade as planets, the same technique could be used to probe the structure of known warm dust populations. For example, the $\sim$1Gyr old star $\eta$ Corvi hosts 
both warm and cool dust components \citep{2005ApJ...620..492W}, with the warm component residing at about 1au \citep{2009A&A...503..265S,2015ApJ...799...42D}. The origin of the warm component is unknown, but suggested to be the outcome of a system-wide recent or ongoing dynamical instability 
\citep{2011LPI....42.2438L}. Recent observations with the LBTI constrained this dust to lie within a projected distance of 1au \citep{2015ApJ...799...42D}, less than the 3au predicted by modelling of the infrared spectrum \citep{2011LPI....42.2438L}, 
with one possible resolution being that the warm dust component is clumpy. The disc-to-star flux ratio at 12$\mu$m is 1.2, roughly two times higher than the detectable limit. Thus, assuming a Solar mass star, the curved lines in the left panel of Figure 1 show the level of astrometric 
signal that would result if 50\% of the warm component were concentrated in a single clump near 1au (though a factor two overestimated, as $\eta$ Corvi is 18pc distant). The astrometric signal would be a few micro-arcseconds for an optical mission, and around 10 mas 
for mid-IR observations. GRAVITY astrometric precision is on the order of a few $\mu$as and could detect such a star's displacement \citep{2014SPIE.9146E..2EL}. In the mid-IR, the astrometric capabilities of IRAC/Spitzer of $\sim$ 20mas could be used to probe these azimuthal structures 
at longer wavelengths \citep{2016AJ....151....9E}. Thus, high precision astrometry provides a possible way to probe the azimuthal structure of warm dust.

The detection limits that were used on Fig.~\ref{figA} were photometric but for specific systems, it is possible to go deeper. Indeed, it might be possible to rule out a dust spot scenario using nulling interferometry (LBTI), which can go a few orders of magnitude deeper \citep{2016arXiv160106866D}.
However, LBTI requires bright stars, so this option would only be available for the nearest stars, which is the case of most interest as this is where astrometric measurements are most sensitive.

Gaia is going to detect at least 20,000 Jupiter-mass planets or larger \citep{2014ApJ...797...14P}. The astrometric effect of such planets is rather large ($>$ 100$\mu$as according to Eq.~\ref{eqas}) and for most stellar types it would mean that if dust was 
mimicking these planets, it would be detectable. It is clearly shown on Fig.~\ref{figA} where a Jupiter-like planet with $\Delta \alpha=100\mu$as lies above the sensitivity line when observed in the optical. We note that there are more bands than WISE 12$\mu$m 
and they could also be used. Hence, a lack of an IR-excess rules out the possibility that an astrometric detection is dust but an IR-excess does not mean the planet is not real, since the dust could be axisymmetric.

Another source of perturbation is the transit of a planet. Indeed, from the ingress to the egress of the transit, the photo-centre of the stellar surface is displaced by an angle $2 \alpha_\star (R_{p}/R_{\star})^3$, where $ \alpha_\star$ is the angular size of the host star and $R_{p}$ ($R_\star$) is the planet (star) radius \citep{Schneider1999}.
It leads to an astrometric effect of 2$\mu$as for a Jupiter transiting a K dwarf at 50 pc. But it cannot mimic a dynamical perturbation by a planet since its duration
is only that of the transit, i.e. a few hours. Nevertheless, similar transits can be useful to investigate the inner structure of discs \citep{Rodriguez2015}. Indeed, if the disc is seen nearly edge on, its inhomogeneities with an orbital period 
of a few years leading to an astrometric effect could be investigated further by their transits. The depth and duration of the transit give the optical depth and size of the inhomogeneity, which can be used to model a pseudo-astrometric 
displacement of the photo-centre. That measurement would help to disentangle a planet-like astrometric effect from a pseudo-astrometric effect due to a disc inhomogeneity.

\section{Conclusion}
Here we have explored an artefact to be taken into account in any future astrometric detection of exoplanets. We find that dust clouds with cometary masses orbiting close to their host stars (within $\sim$ 1 arcsec)
 produce enough flux to mimic an Earth-like planet at 10pc. We have shown that one can expect a large diversity of situations to produce these inhomogeneities, so that, in future observations each case shall require a specific analysis. Astrometric observations around protoplanetary discs can also be affected by dust inhomogeneities 
as we have shown in section \ref{observed}. We suggest that the astrometric signal observed around
VB10 (potentially the first astrometric detection of a planet), which does not present any radial velocity might be explained by a massive dust cloud orbiting VB10 rather than a Jupiter-like planet. Our method can be used for the coming astrometric detection of thousands of planets with Gaia to rule out potential dust clumps.
In any case, the most secure test to confirm a planet detection with astrometry shall be to check for the achromaticity of the astrometric measurement, and radial velocity measurements whenever possible. However, finding systems that are not achromatic would be interesting as it would reveal the detection of a certain amount of dust within the inner parts of planetary systems.
This provides a new technique to probe the unresolved part of planetary systems and check for asymmetries due to dust clouds, exozodis or other inhomogeneities presented in this paper.

\begin{acknowledgements}
QK acknowledges support from the European Union through ERC grant number 279973D. GMK is supported by the Royal Society as a Royal Society University Research Fellow. 
Souami thanks the "Action F\'ed\'eratrice Exoplan\`etes" of the Paris Observatory for financial support. QK wishes to thank 
J.-C. Augereau for fruitful discussions about GRaTeR. JS wishes to thank Didier Queloz for an initial discussion.
\end{acknowledgements}

\end{document}